\documentclass[aps,pre,twocolumn,showpacs]{revtex4}
\usepackage{amsmath}
\begin{document}
\title{Scaling in a general class of critical random Boolean networks}
\author{Tamara Mihaljev and Barbara Drossel}
\affiliation{Institut f\"ur Festk\"orperphysik,  TU Darmstadt,
Hochschulstra\ss e 6, 64289 Darmstadt, Germany }
\date{\today}
\begin{abstract}

We derive analytically the scaling behavior in the thermodynamic limit
of the number of nonfrozen and relevant nodes in the most general
class of critical Kauffman networks for any number of inputs per node,
and for any choice of the probability distribution for the Boolean
functions. By defining and analyzing a stochastic process that
determines the frozen core we can prove that the mean number of
nonfrozen nodes in any critical network with more than one input per
node scales with the network size $N$ as $N^{2/3}$, with only
$N^{1/3}$ nonfrozen nodes having two nonfrozen inputs and the number
of nonfrozen nodes with more than two inputs being finite in the
thermodynamic limit. Using these results we can conclude that the mean
number of relevant nodes increases for large $N$ as $N^{1/3}$, with only a
finite number of relevant nodes having two relevant inputs, and a
vanishing fraction of nodes having more than three of them. It follows
that all relevant components apart from a finite number are simple
loops, and that the mean number and length of attractors increases
faster than any power law with network size.
\end{abstract}
\maketitle
\section{Introduction}
\label{intro}

Random Boolean networks are often used as generic models for the dynamics of
complex systems of interacting entities, such as social and economic
networks, neural networks, and gene or protein interaction networks
\cite{kauffman:random}. The simplest and most widely studied of these
models was introduced in 1969 by Kauffman \cite{kauffman:metabolic} as
a model for gene regulation.  The system consists of $N$ nodes, each
of which receives input from $K$ randomly chosen other nodes. The
network is updated synchronously, the state of a node at time step $t$
being a Boolean function of the states of the $K$ input nodes at the
previous time step, $t-1$.  The Boolean updating functions are
randomly assigned to every node in the network, and together with the
connectivity pattern they define the realization of the network. For
any initial condition, the network eventually settles on a periodic
attractor. 

Of special interest are \emph{critical} networks, which lie at the
boundary between a frozen phase and a chaotic phase
\cite{derrida:random,derrida:phase}.  In the frozen phase, a
perturbation at one node propagates during one time step on an average
to less than one node, and the attractor lengths remain finite in the
limit $N\to \infty$. In the chaotic phase, the difference between two
almost identical states increases exponentially fast, because a
perturbation propagates on an average to more than one node during one
time step \cite{aldana-gonzalez:boolean}. Whether a network is frozen,
chaotic, or critical, depends on the connectivity $K$ as well as on
the weights of the different Boolean functions. If these weights are
chosen appropriately, critical networks can be created for any value
of $K$.

The nodes of a critical network can be classified according to their
dynamics on an attractor. First, there are nodes that are frozen on
the same value on every attractor. Such nodes give a constant input to
other nodes and are otherwise irrelevant. They form the \emph{frozen
core} of the network. Second, there are nodes whose outputs go only to
irrelevant nodes. Though they may fluctuate, they are also classified
as irrelevant since they act only as slaves to the nodes determining
the attractor period. Third, the \emph{ relevant nodes} are the nodes
whose state is not constant and that control at least one relevant
node. These nodes determine completely the number and period of
attractors. If only these nodes and the links between them are
considered, they form loops with possibly additional links and chains
of relevant nodes within and between loops.  The recognition of the
relevant elements as the only elements influencing the asymptotic
dynamics was an important step in understanding the attractors of
Kauffman networks. The behavior of the frozen core was first studied
by Flyvbjerg \cite{flyvbjerg:order}. Then, in an analytical study of
$K=1$ networks Flyvbjerg and Kjaer \cite{flyvbjerg:exact} introduced
the concept of relevant elements.  This concept was generalized to
general critical networks by Bastolla and Parisi
\cite{bastolla:relevant,bastolla:modular}. They gained insight into
the properties of the attractors of the critical networks by using
numerical experiments based on the modular structure of the relevant
elements. Finally, Socolar and Kauffman \cite{socolar:scaling} found
numerically that for critical $K=2$ networks the mean number of
nonfrozen nodes scales as $N_{nf}\sim N^{2/3}$, and the mean number of
relevant nodes scales as $N_{rel}\sim N^{1/3}$. The same result is
hidden in the analytical work on attractor numbers by Samuelsson and
Troein \cite{samuelsson:superpolynomial}, as was shown in
\cite{drossel:onnumber}. An explicit analytical derivation of these
and other scaling laws was given in \cite{kaufmanandco:scaling}.
For $K=1$, these power laws are $N_{nf}\sim N$ and $N_{rel}\sim
N^{1/2}$, since there is no frozen core in a $K=1$ critical network.

In this work, we will derive the scaling behavior of the number of
nonfrozen and of relevant nodes in critical Kauffman networks with $K
\ge 3$. Since the scaling behavior is different for $K=1$ and $K=2$,
one could expect that the exponents are generally
$K$-dependent. However, we will show that the exponents $2/3$ and
$1/3$ found for $K=2$ are valid also for larger $K$ and for all
possible probability distributions of the Boolean functions, as long
as the network is critical.  We also obtain results for the number of
nonfrozen nodes with two and more nonfrozen inputs, and for the number
of relevant nodes with two and more relevant inputs.

The outline of this paper is the following. In the next section, we
introduce a stochastic process that yields the frozen core in $K=3$
networks. The mean-field-theory for this process is presented in
Section \ref{meanfield}, and an improved treatment including
fluctuations is presented in Section \ref{fluctuations}, yielding the
scaling behavior of the number of nonfrozen nodes in critical
networks. The next three sections are devoted to special points in
parameter space, where the stochastic process does not generate all of
the frozen core. In Sections \ref{specialpoints} and
\ref{selffreezing} those points are considered, where the stochastic
process gives a smaller frozen core, and it is shown that
``self-freezing loops'' generate the rest of the frozen core. In
Section \ref{noconstantfunctions}, we consider points in parameter
space, where the stochastic process does not generate any frozen
nodes, and where self-freezing loops are responsible for all of the
frozen core. Finally, in Sections \ref{generalization} and
\ref{relevant} we evaluate the case $K \ge 4$ and the scaling behavior
of the relevant nodes and attractor properties. Section
\ref{conclusions} discusses the implications of our results.


\section{A stochastic process that leads to the frozen core}
\label{process}

From now on, we set $K=3$ and derive explicitly the scaling behavior
of the nonfrozen nodes. The generalization to larger $K$ and the
scaling behavior of the relevant nodes will be discussed later. 
The first step of the calculation, which is performed in this section,
consist in defining a stochastic process that determines the frozen
core. This process is inspired by the one used in
\cite{kaufmanandco:scaling} for $K=2$, however it needed to be
modified before it could be generalized to larger $K$. 
The treatment presented in the following is
based on the existence of nodes with constant functions (functions in
which the output
is fixed irrespectively of the input) and it
therefore applies to all critical models that have a nonzero fraction of
constant functions. Networks with no constant functions, and in
particular networks with only canalyzing functions will be
discussed separately. 

Flyvbjerg \cite{flyvbjerg:order} was the first one to use a dynamical
process that starts from the nodes with constant update functions and
determines iteratively the frozen core. Performing a mean-field
calculation for this process, he could identify the critical point. We
define in the following a process that goes beyond mean-field theory
and gives exact results for the frozen core.  We consider the ensemble
of all networks of size $N$ with a fixed number of nodes with constant
update functions. All nodes with a constant update function are
certainly part of the frozen core. We construct the frozen core by
determining stepwise all those nodes that become frozen due to the
influence of a frozen node. In the language of \cite{socolar:scaling},
this process determines the ``clamped'' nodes.

In a $K=3$ network, each node has 3 inputs, and there are consequently
$2^{2^3}=256$ possible Boolean functions. In order to specify a model,
one has to specify the probabilities for a node to choose each of
these functions. Instead of performing the calculation
in terms of all these parameters, it turns out that three parameters
are sufficient. For the $K=2$ networks, we introduced 3 parameters
corresponding to the occurrence of three types of Boolean
functions. For larger $K$, there are more types of Boolean functions,
and we use therefore a different set of parameters. The first
parameter is $\beta$, which is the proportion of nonfrozen nodes in
the network. $1-\beta$ is therefore the proportion of nodes with a
constant update function. We require $\beta < 1$ for the calculation
performed in this and the following section. The case $\beta = 1$ will
be discussed further below. The second parameter is $\omega_2$, which
is the probability that a randomly chosen node that does not have a
constant update function will become a frozen node when one of its 3
inputs is connected to a frozen node.  If one input of a node is fixed at some
value, the node has effectively two inputs left. We now consider those nodes
that have not become frozen by fixing one input, i.e. we are considering the
proportion $1-\omega_2$ of all nonfrozen nodes. The parameter
$\omega_1$ is then the probability that such a node becomes frozen
when one of the remaining two inputs is connected to a frozen node.
This probability can again be expressed in terms of the probabilities
of the different possible update functions.  Thus all the networks
with the same parameters $\omega_2$, $\omega_1$ and $\beta$ will be
treated as of the same type.  As we will see below, the properties we
are interested in will be the same not only for the functions that
belong to the same type of the network (i.e., that have the same
parameters but possibly different Boolean functions) but also for the
different types as long as their parameters are such that the network
satisfies the criticality condition (\ref{critcond}) derived below.
This means that we can have critical networks with all possible
choices of Boolean functions and that they will all be characterized
by the same exponents as a consequence of being critical.

Now, let us define the stochastic process that determines the frozen
core. For this purpose, we differentiate 4 types of nodes, the numbers
of which will change during the process, and we place these nodes in 4
different ``containers''. Initially, all nodes with constant functions
are placed in a container labelled $\mathcal{F}$, and the
remaining
nodes in a container labelled $\mathcal{N}_3$. In this container are
all those nodes, for which we do not yet know if they are connected to
a frozen node. The other two containers, labelled $\mathcal{N}_2$ and
$\mathcal{N}_1$, are initially empty. They will contain nodes with one
and two frozen inputs that are themselves not (yet) frozen.  Since the
number of nodes in the different containers is going to change during
our stochastic process, we denote the initial values of numbers of
nodes in the containers as $N_f^{ini}$, $N_2^{ini}$=$N_1^{ini}=0$ and
$N_3^{ini}$, and the total number of nodes as $N^{ini}$ (this is the
actual number of nodes in the network).  The contents of the
containers will change with time. The ``time'' we are defining here is
not the real time for the dynamics of the system.  Instead, it is the
time scale for a stochastic process that we use to determine the
frozen core. During one time step, we choose one node from the
container $\mathcal{F}$ and determine the influence of this node on
the nodes connected to it. After determining its influence we will
remove it from the system, and the number of nodes $N$ in the system
is reduced by 1. Now, for each nonfrozen node in container
$\mathcal{N}_3$ we ask whether it receives input from the chosen
frozen node. If this is the case it freezes with probability
$\omega_2$ due to the influence of this node and moves to container
$\mathcal{F}$. With probability $1-\omega_2$ it does not become frozen
and moves to container $\mathcal{N}_2$. In one time step, we therefore
move each node of container $\mathcal{N}_3$ with probability $3
\omega_2 /N$ to the container $\mathcal{F}$, and with probability
$3(1-\omega_2)/N$ to the container $\mathcal{N}_2$. Similarly, a node
from the container $\mathcal{N}_2$ receives input from the chosen
frozen node with probability $2/N$, and it will then become frozen
with probability $\omega_1$ and will be placed in the container
$\mathcal{F}$.  If it does not freeze, we place it in container
$\mathcal{N}_1$, where we find all those nodes that have two inputs
from frozen nodes and are not frozen. When nodes from this container
choose a frozen node as an input, they automaticly become
frozen. During this process, the probabilities $\omega_2$ and
$\omega_1$ will not change since the nodes from containers
$\mathcal{N}_3$ and $\mathcal{N}_2$, for which we are in every time
step determining whether they are going to freeze, are chosen at
random, and moving them from the containers will not change
probability distribution of the functions of the nodes left in the
containers.  In the next time step, we choose another frozen node from
container $\mathcal{F}$ and determine its effect on the other
nodes. Some nodes move again to a different container, and the chosen
frozen node is removed from the system. We repeat this procedure until
we can not continue because either container $\mathcal{F}$ is empty, or
because all the other containers are empty. If container $\mathcal{F}$
becomes empty, we are left with the nonfrozen nodes. We shall see
below that most of the remaining nodes are in container
$\mathcal{N}_1$, with the proportion of nodes left in containers
$\mathcal{N}_2$ and $\mathcal{N}_3$ vanishing in the limit $N^{ini}\to
\infty$. If all containers apart from container $\mathcal{F}$ are
empty at the end, the entire network becomes frozen. This means that
the dynamics of the network goes to the same fixed point for all
initial conditions.
\section{Mean field approximation and the criticality condition}
\label{meanfield}

Let us first describe this process by deterministic equations that
neglect fluctuations around the average change of the number of nodes
in the different containers.
As long as all containers contain large numbers of nodes, these
fluctuations are negligible, and the deterministic description is
appropriate. The average change of the node numbers in the containers
during one time step is
\begin{eqnarray}
\Delta N_3 &=& - \frac{3N_3}{N}\nonumber \\
\Delta N_2 &=& - \frac{2N_2}{N}+(1-\omega_2)\frac{3N_3}{N}\nonumber \\
\Delta N_1 &=& - \frac{N_1}{N} + (1-\omega_1)\frac{2N_2}{N}  \label{Delta}\\
\Delta N_f &=& -1 +  \frac{N_1}{N} + \omega_1\frac{2N_2}{N}+\omega_2\frac{3N_3}{N}\nonumber \\
\Delta N &=& -1\nonumber
\end{eqnarray}
The total number of nodes in the containers, $N$, can be used instead of the
time variable, since it decreases by one during each step. The
equation for $N_3$ can then be solved by going from a difference
equation to a differential equation,
$$\frac {\Delta N_3}{\Delta N} \simeq \frac {d N_3}{d N} =  -
\frac{3N_3}{N}\, ,$$
which has the solution
\begin{equation}
N_3 = N^3 \frac{N_3^{ini}}{(N^{ini})^3} = \frac{\beta}{(N^{ini})^2}N^3\, ,\nonumber\\
\end{equation}
where $\beta=\frac{N_3^{ini}}{N^{ini}}$.
Similarly, we find 
\begin{eqnarray}
N_2 &=&3(1-\omega_2)\frac{\beta}{N^{ini}}N^2 - 3(1-\omega_2)\frac{\beta}{(N^{ini})^2}N^3\nonumber\\
N_1 &=& 3(1-\omega_1)(1-\omega_2)\beta N - 6(1-\omega_1)(1-\omega_2)
\frac{\beta}{N^{ini}}N^2
\nonumber\\ && +3(1-\omega_1)(1-\omega_2)\frac{\beta}{(N^{ini})^2}N^3\nonumber
\label{det}\\
N_f &= &(1-3(1-\omega_1)(1-\omega_2)\beta)N \nonumber\\ &&+
3(1-2\omega_1)(1-\omega_2)\frac{\beta}{N^{ini}}N^2  \nonumber\\ &&+
(3\omega_1(1-\omega_2)-1)\frac{\beta}{(N^{ini})^2} N^3 
\end{eqnarray}

When $1-3(1-\omega_1)(1-\omega_2)\beta<0$, the equation $N_f=0$,
which represents the stopping condition for the process, has a
solution for an nonzero value $N$. This solution shows that the number of
nonfrozen nodes in each container is proportional to $N^{ini}$. This
means that on an average a nonfrozen node has more than one nonfrozen
input. A perturbation at one node propagates during one time step on
an average to more than one node and we are obviously in the chaotic
phase.  

For $1-3(1-\omega_1)(1-\omega_2)\beta\geq 0$ the equation $N_f=0$ does
not have a nonzero solution for $N\in [0,N^{ini}]$. In this case,
we will stop the process when $N_f$ drops below 1.  We are in the frozen
phase, or we have a  critical system. 

In the case $1-3(1-\omega_1)(1-\omega_2)\beta>0$, the values $N_3$ and
$N_2$ will sink below 1 when $N$ becomes of the order
$\sqrt{N^{ini}}$, and the higher-order terms contributing to $N_f$ and
$N_1$ can be neglected compared to the first one. For smaller $N$,
only frozen nodes and nodes with one input are left. When $N_f$ falls below
1, there remain only a constant number of the nodes of type
$\mathcal{N}_1$, $$N_1\simeq \frac{3(1-\omega_1)(1-\omega_2)\beta}{
1-3(1-\omega_1)(1-\omega_2)\beta }\, .$$ The network is essentially
frozen, with only a finite number of nonfrozen nodes in the limit
$N^{ini} \to\infty$. If we now choose the inputs for these nodes, we
obtain simple loops with trees rooted in the loops. This property of
the frozen phase was also found in \cite{socolar:scaling}.
 
When parameters of the networks are such that
\begin{equation}
1-3(1-\omega_1)(1-\omega_2)\beta=0 \label{critcond}
\end{equation} 
is fulfilled, we are at the boundary
between frozen and chaotic phase in the parameter space. Thus the
network is critical. 
Since the stochastic process stops at $N_f = 1$, we have
\begin{equation*}
1=\frac{(1-2\omega_1)}{(1-\omega_1)}\frac{(N^{end})^2 }{N^{ini}} +
\left(\frac{\omega_1}{(1-\omega_1)}-\beta\right)\frac{(N^{end})^3}{
(N^{ini})^2}   \, .
\end{equation*} 
In the limit  $N^{ini} \to\infty$ the first term is dominant and the
number of nonfrozen nodes would scale with the square root of the
network size if the deterministic approximation to the stochastic
process was exact. We shall see below that including fluctuations
changes the exponent from $1/2$ to $2/3$.  The final number of
$\mathcal{N}_2$-nodes for the deterministic process for the critical
networks  is independent of network size, and
the final number of $\mathcal{N}_3$-nodes is $\sim (N^{ini})^{-1/2}$
and vanishes for  $N^{ini} \to\infty$ .
 We shall see below that the fluctuations change these two
results to $N_2\sim(N^{ini})^{1/3}$ and $N_3\sim const$.  

The deterministic description of our process gives the wrong scaling
of the number of nonfrozen nodes in the case of critical networks, but
a correct criticality condition (\ref{critcond}).  We are interested
in the dynamical behavior of the networks in the critical phase and we
will from now on study only networks with the parameters such that the
criticality condition $1-3(1-\omega_1)(1-\omega_2)\beta=0$ is
fulfilled.

Before we proceed by introducing the noise into the deterministic
equations, there is one more piece of information we can extract from
the deterministic description of the critical process that is going to
help us later in determining the noise term.  Introducing $n =
N/N^{ini}$ and $n_j = N_j/N^{ini}$ for $j=f,1,2,3$, equations
(\ref{det}) simplify to (using the criticality condition)
\begin{eqnarray}
n_3 &=& \beta n^3\nonumber\\
n_2 &=& \frac{1}{1-\omega_1}(n^2-n^3) \nonumber\\
n_1 &=& n-2n^2 + n^3 \nonumber\\
n_f &=&  \frac{1-2\omega_1}{1-\omega_1} n^2 + \left(\frac{\omega_1}{1-\omega_1}-\beta\right)n^3
\, .
\end{eqnarray}
This means that our stochastic process remains invariant (in the
deterministic approximation) when the initial number of nodes in the
containers and the time unit are all multiplied by the same factor.
For small $n$, the majority of nodes are in container $\mathcal N_1$,
since $n_1= n - \mathcal{O}(n^2)$. Now, if we choose a
sufficiently large $N^{ini}$, $n$ reaches any given small value while
$N_f\sim n^2 N^{ini} $ is still large enough for a deterministic
description.  We can therefore assume that for sufficiently large
networks $N_f/N\sim n$ becomes small before the effect of the noise
becomes important. This assumption will simplify our calculations
below.

\section{The effect of fluctuations}
\label{fluctuations}

The number of nodes in container $\mathcal{N}_j$ , $j=1,2,3\,$, that
choose a
given frozen node as an input is Poisson distributed with a mean $jN_j/N$
and a variance $jN_j/N$. We now assume that $n$ is small at the
moment where the noise becomes important, i.e., that the variance of the
three noise terms is $N_1/N = n_1/n = 1-2n+n^2 = 1-\mathcal{O}(n)$
 and $2N_2/N =2n_2/n = \frac{2}{1-\omega_1}(n-n^2) =
\mathcal{O}(n)$ and $3N_3/N = 3\beta n^2 = \mathcal{O}(n^2)$. 
All three noise terms occur in the equation for $N_f$, and since the
first term dominates for small $n$, we consider only this term in the
equation for $N_f$. In the equations for $N_1$ and $N_2$, the noise
term is much smaller than the number of nodes in these containers and
can therefore be dropped. 

The effect of the noise on the final value of $N_3$ can be obtained by
the following consideration: as we will see below, the mean final
value of $N_3$ will be a constant, which is independent of
$N^{ini}$. This means that each node that is initially in the
container $\mathcal{N}_3$ has a probability of the order $1/N^{ini}$
of never choosing a frozen input during the stochastic process, and
this probability is independent for each node. From this follows that
the final number $N_3$ is Poisson distributed with a variance that is
identical to the mean. This variance is finite in the limit
$N^{ini}\to\infty$ and it does not affect the final value of $N_2$ or
$N_1$. Since we have obtained the variance of the final value of $N_3$
by this simple argument, we will not explicitly consider the noise
term in the equation for $N_3$. 

We therefore obtain the stochastic version of equations (\ref{Delta}),
where we need to retain only the noise term in the equation for $N_f$:
\begin{eqnarray}
\Delta N_3 &=& - \frac{3N_3}{N}\nonumber \\
\Delta N_2 &=& - \frac{2N_2}{N}+\frac{1}{\beta(1-\omega_1)}\frac{N_3}{N}\nonumber \\
\Delta N_f &=& -1 +  \frac{N_1}{N} + 2\omega_1\frac{N_2}{N}+\left(3-\frac{1}{\beta(1-\omega_1)}\right)\frac{N_3}{N}-\xi \nonumber \\
\Delta N &=& -1\, .\end{eqnarray}
The random variable $\xi$ has zero mean and unit variance. As long as
the $n_j$ change little during one time step, we can summarize a large
number $T$ of time steps into one effective time step, with the noise
becoming Gaussian distributed with zero mean and variance $T$. Exactly
the same process would result if we summarized $T$ time steps of a
process with Gaussian noise of unit variance. For this reason, we can
choose the random variable $\xi$ to be Gaussian distributed with unit
variance.

Compared to the deterministic case, the equations for $N_3$ and
$N_2$ are unchanged. Inserting the solution for $N_3$ and $N_2$ into
the equation for $N_f$, we obtain
\begin{equation}
\frac{dN_f}{dN} = \frac{N_{f}}{N} + \frac{1-2\omega_1}{1-\omega_1}\frac{N}{N^{ini}}+2\left(\frac{\omega_1}{1-\omega_1}-\beta\right)\left(\frac{N}{N^{ini}}\right)^2   
+\xi\label{langevin}
\end{equation}
with the step size $dN=1$ and $\langle \xi^2\rangle = 1$. (In the
continuum limit $dN\to 0$ the noise correlation becomes $\langle
\xi(N)\xi(N')\rangle = \delta(N-N')$).  This is a Langevin-equation,
and the corresponding Fokker-Planck-equation is
\begin{eqnarray}
-\frac{\partial P}{\partial N} &=& \frac{\partial}{\partial
    N_f} \left[ \frac{N_{f}}{N} +
    \frac{1-2\omega_1}{1-\omega_1}\frac{N}{N^{ini}}\right.\nonumber\\
    &+&\left. 2\left(\frac{\omega_1}{1-\omega_1}-\beta\right)\left(\frac{N}{N^{ini}}\right)^2
    \right] P \nonumber\\ &+& \frac 1 2 \frac{\partial^2 P}{\partial
   N_f^2}\, . \label{FP}
\end{eqnarray}

Since we are investigating networks in the thermodynamic limit,
keeping only the leading terms will give a good approximation. Thus,
we can neglect the last term in the expression under the partial
derivative with respect to $N_f$ once  $N/N^{ini}$ has become
sufficiently small. We are left with the Fokker-Planck
equation of the same type as the one already studied in
\cite{kaufmanandco:scaling}, but with a different coefficient. 
\begin{equation}
-\frac{\partial P}{\partial N} = \frac{\partial}{\partial
    N_f} \left( \frac{N_{f}}{N} +  \frac{\mu
  N}{N^{ini}} \right)P + \frac 1 2 \frac{\partial^2 P}{\partial
   N_f^2}\, ,
\end{equation}
where $\mu=(1-2\omega_1)/(1-\omega_1)$\, .

We introduce the variables 
\begin{equation}
x = \frac {N_f}{\sqrt{N}} \hbox{ and } y = \frac{N}{(N^{ini}/\mu)^{2/3}}\label{defxy}
\end{equation}
and the function $f(x,y)= (N^{ini}/\mu)^\gamma P(N_f,N)$. The free
parameter $\gamma$ will be fixed below by the condition that the
probability distribution of the number of nonfrozen nodes is
normalized.  The Fokker-Planck equation then becomes
\begin{equation}
y\frac{\partial f}{\partial y} +f+\left(\frac x 2 + y^{3/2}\right)
    \frac{\partial f}{\partial
    x} + \frac 1 2 \frac{\partial^2 f}{\partial
   x^2} = 0\, . \label{FP2}
\end{equation}
Let $W(N)$ denote the probability that $N$ nodes are left at the
moment where $N_f$ reaches the value zero. It is
\begin{eqnarray*}
W(N)  &=& \int_0^\infty P(N_f,N)dN_f -  \int_0^\infty
P(N_f,N-1)dN_f 
\end{eqnarray*}
Consequently,
\begin{eqnarray*}
W(N) &=& \frac{\partial}{\partial N}\int_0^\infty P(N_f,N)dN_f\nonumber\\
&=&  (N^{ini}/\mu)^{-\gamma -1/3} \frac{\partial}{\partial y}\sqrt y \int_0^\infty f(x,y)dx
\nonumber\\
&\equiv&  (N^{ini}/\mu)^{-\gamma -1/3}  G(y)\label{w}
\end{eqnarray*}
with a scaling function $G(y)$. $W(N)$ must be a normalized function,
$$\int_0^\infty W(N)dN =(N^{ini}/\mu)^{-\gamma
-1/3+2/3}\int_0^\infty G(y)dy = 1 \, .$$ 
This gives $\gamma=1/3$.  This condition is independent of the
parameters of the model, and therefore $G(y)$ and $f(x,y)$ are
independent of them, too. Now, we have
\begin{eqnarray*}
W(N)=(N^{ini}/\mu)^{-2/3}  G(y)
\end{eqnarray*}

The mean number of nonfrozen nodes is
\begin{equation*}\bar N = \int_0^\infty N W(N)dN =  (N^{ini}/\mu)^{2/3} 
\int_0^\infty G(y)
ydy\, ,
\end{equation*}
which is proportional to $ (N^{ini})^{2/3}$. 

The probability $W_2(N_2)$ that $N_2$ nodes are left in container 
 $\mathcal{N}_2$ at the moment where container $\mathcal{F}$ becomes
 empty, is obtained from the relation $$N_2=\frac{1}{1-\omega_1}\frac{N^2}{N^{ini}} - \frac{1}{1-\omega_1}\frac{N^3}{(N^{ini})^2}\nonumber \, .$$
 
Since $W(N)dN = W_2(N_2)dN_2$, we find that
the mean number of nodes left in container 
 $\mathcal{N}_2$ is
\begin{eqnarray*}\bar N_2 &=& \int_0^\infty N_2W_2(N_2)dN_2
=\int_0^\infty N_2 W(N)dN \\
&=&\frac{1}{(\mu)^{1/3}(1-2\omega_1)}(N^{ini})^{1/3} \int_0^\infty
y^2G(y)dy \\ && + \frac{1}{\mu } \int_0^\infty y^2G(y)dy\sim(N^{ini})^{1/3}\, .
\end{eqnarray*}
In the same manner we find for the number of nodes left in container $\mathcal{N}_3$
\begin{eqnarray*}\bar N_3 &=& \int_0^\infty N_3 W_3(N_3)dN_3 =\int_0^\infty N_3
W(N)dN
\\ &=&\frac{\beta (1-2\omega_1)^2}{(1-\omega_1)^2} \int_0^\infty y^3G(y)dy \sim const\, .
\end{eqnarray*}

Thus, we have shown that the number of nonfrozen nodes scales with
network size $N^{ini}$ as $(N^{ini})^{2/3}$, with most of these nodes
receiving only one input from other nonfrozen nodes. The number of
nonfrozen nodes with two nonfrozen inputs scales as $(N^{ini})^{1/3}$
and the number of nodes with three such inputs is independent of the
network size.

\section{Special points and canalyzing functions}
\label{specialpoints}

For $\omega_1 = 1/2$, the second term in the
Langevin Equation (\ref{langevin}) is zero. In this case the next
order term has to be taken into account since it is the leading one
now. We will see that the  mechanism of creating the frozen core
is different for such systems, but  in the end we will find
the same scaling behavior of the number of nonfrozen nodes.

Now we have to consider the modified
Langevin equation
\begin{equation}
\frac{dN_f}{dN} = \frac{N_{f}}{N} +2(1-\beta)\left(\frac{N}{N^{ini}}\right)^2   
+\xi\label{langevin2}
\end{equation}
and the corresponding Fokker-Planck equation
\begin{equation}
-\frac{\partial P}{\partial N} = \frac{\partial}{\partial
    N_f} \left( \frac{N_{f}}{N} +2(1-\beta)\left(\frac{N}{N^{ini}}\right)^2 \right)P + \frac 1 2 \frac{\partial^2 P}{\partial
   N_f^2}\, . \label{FP1}
\end{equation}

We again introduce new variables
\begin{equation}
x = \frac {N_f}{\sqrt{N}} \hbox{ and } y = \left(\frac{(N^{ini})^2}{2(1-\beta)}\right)^{-4/5}N^2
\end{equation}
and the function $f(x,y)=
\left(\frac{(N^{ini})^2}{2(1-\beta)}\right)^\gamma P(N_f,N)$.
The Fokker-Planck equation then becomes
\begin{equation*}
2y\frac{\partial f}{\partial y} +f+\left(\frac x 2 + y^{5/4}\right)
    \frac{\partial f}{\partial
    x} + \frac 1 2 \frac{\partial^2 f}{\partial
   x^2} = 0\, . 
\end{equation*}

For the probability that $N$ nodes are left when $N_f$ reaches zero we
obtain
\begin{eqnarray*}
W(N)=\left(\frac{(N^{ini})^2}{2(1-\beta)}\right)^{-2/5} \tilde G(y)
\end{eqnarray*}
 with a new scaling function $\tilde G$. We have used the fact that this probability has to be
normalized, which gives $\gamma =1/5$.

 Using this result, we find for the mean number of nonfrozen nodes 

\begin{eqnarray}\bar N &=& \int_0^\infty N W(N)dN = \frac{1}{2}
\left(\frac{(N^{ini})^2}{2(1-\beta)}\right)^{2/5} \int_0^\infty
\tilde G(y)dy\nonumber\\ &\sim & (N^{ini})^{4/5}\, .\label{barN}
\end{eqnarray}
 
 For the mean number of nonfrozen nodes left in
containers $\mathcal{N}_2$ and $\mathcal{N}_3$ we find
 
\begin{eqnarray}\bar N_2 &=& \int_0^\infty N_2W_2(N_2)dN_2
=\int_0^\infty N_2 W(N)dN
\nonumber \\ &=&\frac{(N^{ini})^{3/5}}{(2(1-\beta))^{4/5}}\int_0^\infty y^{1/2}\tilde G(y)dy
\nonumber\\ && - \frac{(N^{ini})^{2/5}}{(2(1-\beta))^{6/5}} \int_0^\infty
y\tilde G(y)dy\nonumber\\ &\sim & (N^{ini})^{3/5}\label{barN2} 
\end{eqnarray}
 and 
\begin{eqnarray}\bar N_3 &=& \int_0^\infty N_3W_3(N_3)dN_3
=\int_0^\infty N_3W(N)dN \nonumber\\ &=& \frac{\beta}{2}
\frac{(N^{ini})^{2/5}}{(2(1-\beta))^{6/5}} \int_0^\infty
y\tilde G(y)dy\nonumber\\ &\sim & (N^{ini})^{2/5}\, .\label{barN3} 
\end{eqnarray}

We see that the number of nodes which become frozen due to the
influence of the constant functions is smaller than in the case of
other
critical networks. When we look at the parameters for these networks
more closely, we see that these networks are effectively canalyzing
with two inputs per node. The probability that a node with two inputs
is going to freeze during one time step is $\omega_1=1/2$ and this
means that the network has Boolean functions such that nodes with two
nonfrozen inputs effectively belong to the ${\mathcal{C}}_1$ or
${\mathcal{C}}_2$ class of Boolean functions with two variables,
i.e., canalyzing functions. The class ${\mathcal{C}}_1$
contains those functions that depend only on one of the two variables,
but not on the other one. The class ${\mathcal{C}}_2$ contains the
remaining canalyzing functions, where one state of each input fixes
the output.
It has been shown in \cite{ute:canal} that in $K=2$
networks with only this type of functions another mechanism of
creating the frozen core arises. The only condition for this is that
the number of nodes from class ${\mathcal{C}}_2$ is large enough. We
will show that it is exactly what happens in the networks we are
analyzing now. The number of nonfrozen nodes with two inputs and
canalyzing ${\mathcal{C}}_2$ functions is here large enough to allow
for the creation of the self-freezing loops that are going to increase
the number of frozen nodes and thus change the scaling of the
nonfrozen nodes from $(N^{ini})^{4/5}$ to $(N^{ini})^{2/3}$.
\section{Creating self-freezing loops and their effect}
\label{selffreezing}

We are now considering a reduced network consisting of those nodes
that are not frozen through the influence of the nodes with constant
functions. The size of this network is $N\simeq (N^{ini})^{4/5}$, most
of the nodes have one nonfrozen input, $N_2\simeq (N^{ini})^{3/5}$
have two, and $N_3\simeq (N^{ini})^{2/5}$ have three nonfrozen inputs.
Nodes with two nonfrozen inputs have a probability to freeze
$\omega_1=1/2$ and as such effectively have canalyzing Boolean
functions of two arguments, belonging to ${\mathcal{C}}_1$ or
${\mathcal{C}}_2$ class. So, the number of nodes with two nonfrozen
inputs that belong to the ${\mathcal{C}}_2$ class has to be $\simeq
(N^{ini})^{3/5}$ as it is the fraction of all nonfrozen nodes with two
inputs.

Let us now assume that there exist groups of nodes that fix
each other's value and do not respond to changes in nodes outside this
group. The simplest example of such a group is a loop of
${\mathcal{C}}_2$ nodes where each node canalyzes (fixes) the state of
its successor once it settles on its majority bit (the one occurring 3
times in its update function table).
These loops, introduced in \cite{ute:canal}, are called
\emph{self-freezing loops}. They can also
contain chains of nodes with one nonfrozen input or with
two nonfrozen inputs and a ${\mathcal{C}}_1$ function between
${\mathcal{C}}_2$ nodes.
If a chain between two ${\mathcal{C}}_2$ nodes as a whole inverts the
state of the first ${\mathcal{C}}_2$ node, the
inverted majority bit of the first ${\mathcal{C}}_2$ node has to canalyze the
second ${\mathcal{C}}_2$ node. The only effect of nodes with
${\mathcal{C}}_1$ functions and those with one nonfrozen input in such loops is
to delay the signal propagation between two adjacent ${\mathcal{C}}_2$ nodes. 
The procedure of finding self-freezing loops is explained in details
in \cite{ute:canal}. The number of nodes on self-freezing loops is
there found by mapping the problem of finding a self-freezing loop
in a ${\mathcal{C}}_2$ network onto the problem of finding the relevant nodes
sitting on relevant loops in a critical network that contains no canalyzing
functions at all, but only reversible (where the output is changed
whenever one of the inputs is changed) and constant functions. Using
results for these reversible networks obtained in
\cite{kaufmanandco:scaling} it was found that the
number of nodes on self freezing loops scales as $\sim N^{1/3}$ where
$N$ is the number of ${\mathcal{C}}_2$ nodes.

Obviously, nodes depending on or canalyzed by the frozen nodes of the
self-freezing loops freeze also, and such nodes may lead to the
freezing of further nodes, etc. We can introduce a dynamical process in
order to determine the total number of nodes that become frozen
because of the self-freezing loops. This process is almost the same as
the one we have used for identifying the influence of the constant
functions on the networks dynamics. We again have
four containers where the nodes left after determining the influence
of the nodes with constant functions are placed. Initially nodes found
to be on the self-freezing loops are going to be moved from the
container with nodes with two inputs, $\mathcal{N}_2$, to the container
$\mathcal{F}$. Thus the initial number of nodes in the containers is
going to be
$N_f^0=((N^{ini})^{3/5})^{1/3}=(N^{ini})^{1/5}$,
$N_2^0=(N^{ini})^{3/5}-N_f^0\simeq (N^{ini})^{3/5}$ and
$N_3^0=(N^{ini})^{2/5}$, and the total number of nodes is
$N^0=(N^{ini})^{4/5}$. Now we run the same dynamical process as
before determining influence of the nodes from the
frozen loops on the rest of this reduced network one by one and then removing
them from the system. At the end of this process we will again have nodes in the container
$\mathcal{N}_2$.
They can now make new self-freezing loops made of ${\mathcal{C}}_2$
nodes with the chains of nodes with one nonfrozen input between them.
We can then again move $N_2^{1/3}$ nodes that are on the new
self-freezing loops to the container $\mathcal{F}$ and run the same
process again. We can even take over the values of
$N_1$, $N_2$ and $N_3$ and $N$ at the end of the first process, since
${N_2}^{1/3}$ frozen nodes moved from container ${\mathcal{N}}_2$ are
negligible in comparison to $N_2$. These processes can be repeated
as long as the number of nodes of type ${\mathcal{C}}_2$ is large
enough to allow for the creation of self-freezing loops. 
The equations for the change of $N_3$ and $N_2$ nodes  
\begin{eqnarray}
\Delta N_3 &=& - \frac{3N_3}{N}\nonumber \\
\Delta N_2 &=& -
\frac{2N_2}{N}+\frac{2}{\beta}\frac{N_3}{N} \label{deltan3n2}
\end{eqnarray}
apply together to all the successive processes of freezing the network through 
the influence of  nodes of the self-freezing loops.
Between each two of them the new self-freezing loops have been found and moved
from the container with $N_2$ nodes allowing for the new
process to start. The equation for $N$ is
$\Delta N= -1$, as before. The solution of these equations is obtained
by going to differential equations for $dN_2/dN$ and $dN_3/dN$. Using the
values of $N$, $N_2$ and  $N_3$, found in Equations (\ref{barN}),
(\ref{barN2}) and
(\ref{barN3}), as initial values of the variables, these differential
equations have the solution  
\begin{eqnarray}
N_3 &=& \frac{N_3^0}{(N^0)^3}N^3\,\label{N3canal} \\
N_2 &=& \frac{N_2^0 + (2/\beta)N_3^0}{(N^0)^2}N^2-\frac{2N_3^0}{\beta(N^0)^3}N^3 \, . \label{N2canal}
\end{eqnarray}
The number of remaining ${\mathcal{N}}_1$ nodes increases in the second
process, the number of ${\mathcal{C}}_2$ (those in container
${\mathcal{N}}_2$)
nodes decreases, thus leading to an increasing weight of ${\mathcal{N}}_1$ nodes
in the nonfrozen network.

The repeated process of identifying generalized self-freezing loops
and the nodes frozen by them breaks down when the remaining nonfrozen
nodes cannot be considered as an effective ${\mathcal{C}}_2$ network any
more. This happens when in the process of creating self-freezing loops the
probability that a ${\mathcal{C}}_2$ node
is going to be attached to the end of the chain of nodes with one
nonfrozen input (thus making closing self-freezing loop possible)
becomes of the same order of magnitude as the probability that this
chain becomes a loop. Since the mean size of the loops of
nodes with one input is found to be of the order of $\sqrt{N}$
\cite{drossel:number} the assembly of self-freezing loop becomes
improbable when $N_2\sim \sqrt{N}$.

This condition gives to leading order
\begin{equation}
\frac{(N^0)^2}{N_2^0} \sim N^{3/2}
\end{equation}
or $N \sim (N^{ini})^{2/3}$.  We again have the same scaling of the
number of nonfrozen nodes with the network size. The scaling of the
number of nonfrozen nodes with two and three nonfrozen inputs with the
network size we find from (\ref{N2canal}) and (\ref{N3canal}) to be
$N_2\sim(N^{ini})^{1/3}$ and $N_3\sim \it{const}$. This is the same
scaling we have for the case of all other
critical networks  investigated until now.

When finding the number of nodes on the self-freezing loops and
defining our second process we assumed that there the influence of the
nodes with three nonfrozen inputs per node is negligible. We can check
if our assumption was justified. In the beginning of this process the
number of nodes with three inputs was $N_3^0\simeq
(N^{ini})^{2/5}$. The number of nodes that are initially on
self-freezing loops is $(N_2^0)^{1/3}=(N^{ini})^{1/5}$. The mean number of
nodes with three inputs on the self-freezing loops is then
$${(N_2^0)^{1/3}}\frac{N_3^0}{N_2^0} = const\, .$$ In the limit of
large network size, only a few (if any) self-freezing loops are destroyed by nodes with three nonfrozen inputs, and this does not change the scaling behavior of the number of nodes on self-freezing loops. 

\section{Networks without constant functions}
\label{noconstantfunctions}

\subsection{Case $\omega_1=1/2$, $\omega_2=1/3$ }

Until now, we have assumed that the network has nodes with constant
functions.  In this section, we consider networks without constant
functions, i.e., with $\beta=1$. The criticality condition
(\ref{critcond}) then becomes $$3(1-\omega_1)(1-\omega_2)=1\, .$$
Although the criticality condition was derived under the assumption
that the network has a nonvanishing proportion of frozen nodes (i.e.,
that $\beta < 1$), it can be extended to $\beta =1$, since it is valid
for any $\beta$ arbitrarily close to 1. Furthermore, decreasing
$\beta$ slightly for fixed $\omega_1$ and $\omega_2$ moves the system
to the frozen phase, indicating that a system satisfying the
criticality condition with $\beta=1$ is at the boundary of the frozen
phase.  As we will see, the value of the parameters in the critical
networks without constant functions we are considering here is
allowing
the formation of the self-freezing loops and leads to the frozen core
of the same size as for all the other critical networks.  Canalyzing
networks and threshold networks are examples of this category of
networks, and they are considered important for biological
applications.

The procedure of creating self-freezing loops in the case of networks
with nodes with two nonfrozen inputs was introduced and explained in
details in \cite{ute:canal}. It is the same procedure we have used in
the previous section. Using a similar line of arguments we can
explain the assembly of the self-freezing loops for the networks with
three inputs per node determined with parameters being $\omega_1=1/2$,
$\omega_2=1/3$ and $\beta=1$. In this case there is a mapping of the
problem of finding the nodes on the self-freezing loops in this
network onto the problem of finding the relevant nodes on relevant
loops in critical network with three inputs per node and only
reversible and constant functions, i.e., with $\omega_1=\omega_2=0$
and
$\beta=1/3$. Self-freezing loops are found by starting with a node and
keeping track of the connection to those inputs that are able to
canalyze this node if they are canalyzed themselves. This procedure is
iterated for these input nodes etc., until a loop is formed or until
it has to stop because no canalyzing inputs are found. Similarly,
relevant loops in a critical network with $\omega_1=\omega_2=0$ are
found by starting with a node and keeping track of the connection to
those inputs that do not have a constant function. This procedure is
iterated for the nonfrozen inputs etc., until a loop is formed or
until it has to stop because no nonfrozen inputs are found. In both
cases, a connection to an input is made with probability 1/3, showing
that the two processes can be mapped on each other.  As we will
show in section $9$ below,  in critical networks with three inputs per node and
nonzero fraction of frozen nodes the number of relevant nodes on relevant
loops scales as $(N^{ini})^{1/3}$.  Therefore, we conclude that in the
network with $\omega_1=\omega_2=0$, the number of nodes on
self-freezing loops scales also as $(N^{ini})^{1/3}$.

We can now proceed just as in the previous section, but with $\beta=1$
and $N_j^0 = N_j^{ini}$. We continue making self-freezing loops and determining which nodes are frozen by them, until  $N_2\sim \sqrt{N}$. Inserting this condition in Equation (\ref{N2canal}), we find to leading order
$$2\frac{N^{3/2}}{N^{ini}} =1\, , $$ leading again to $N \sim
(N^{ini})^{2/3}$.


\subsection{General case}

Now, let us turn to the case $\beta=1$ with $\omega_1 < 1/2$. (The
situation $\omega_1 > 1/2$ is not possible for nonfrozen Boolean
functions with two inputs.)  The probability that a node we don't know
anything about freezes when connected to a frozen node is now
$\omega_2>1/3$. Every node has three inputs and this frozen node could
be any of them. This means that on an average a node can be frozen by
more than one input, and the self-freezing components we look for in
the network here consist of at least as many nodes as those in the
previous subsection. However, we do not need to know the exact number of
frozen nodes in these components. We will build only one self-freezing
loop and move its $(N^{ini})^{1/3}$ nodes to the container
$\mathcal{F}$.  Then we start the calculation of Section \ref{process}
by setting $\beta = 1 - (N^{ini})^{-2/3}$. Since $\omega_1 < 1/2$, the
leading-order terms of the calculation performed in Section
\ref{process} are retained in this case, and we can take over all the
main results of that section. In particular, it follows that a single
self-freezing loop is sufficient to generate the entire frozen core,
and we do not need to identify other self-freezing loops. As before,
the number of nonfrozen nodes scales as $(N^{ini})^{2/3}$.

 
\section{Generalization to larger $K$}
\label{generalization}

The process introduced in Section \ref{process} can easily be
generalized to networks with $K>3$. We first consider again the case
$\beta<1$. For network with $K$ inputs we define a set of parameters
$\beta$ and $\omega_i$ with $i\in [1,K-1]$.  $\beta$ is again
fraction of the nonfrozen nodes and $\omega_i$ is the probability that
a nonfrozen node that has $K-i$ inputs from frozen nodes freezes when
receiving another frozen input in our process.  These K parameters are
going to define completely the class of networks we observe in the
process. Using the deterministic description of the process analogous
to the one described in Section \ref{meanfield} we find the
criticality condition for networks with any $K$:
\begin{equation}
K(1-\omega_1)(1-\omega_2)\cdots (1-\omega_{K-1})=1\, .\label{gencritcond}
\end{equation}
Introduction of noise in  the process gives the
Langevin equation 
\begin{equation}
\frac{dN_f}{dN} =
\frac{N_{f}}{N} +
\sum_{i=1}^{K-1}f_i(\omega_1,\dots,\omega_i)\left(\frac{N}{N^{ini}}
\right)^i 
+\xi
\label{generallangevin}
\end{equation}
where the $f_i(\omega_1,\dots,\omega_i)$ are functions of the
parameters of the system obtained from the stochastic process
They satisfy 
$f_i(\omega_1,\dots,\omega_i)=0$ when $\omega_j =1/(j+1)$ for all
$j\in[1,i]$.
 We see that in this general Langevin equation
the leading term in $N$ is the same as in
Equation (\ref{langevin}).
Therefore we find that in the thermodynamic limit the number of
nonfrozen nodes scales in critical networks as $(N^{ini})^{2/3}$ with
the network size.

Just like in the $K=3$ networks, parameter values can be such that one
or more of the leading terms in the Langevin equation vanish. These
special points in the parameter space describe networks where the
Boolean functions are such that the nodes left nonfrozen after
determining the influence of the frozen nodes in our process can
additionally generate self-freezing loops. Their influence on the rest
of the network has to be determined by generalization of the process
introduced in Section \ref{selffreezing}.  The number of classes of
special points will increase with $K$, leading to a hierarchy
of special points. For each $K$, there are $K-3$ classes of points in
parameter space that are equivalent to the special points of networks
with $K-1$ inputs per node (that is they have the same leading term in
the Langevin equation), and one new class of special points where only
the last term in the Langevin equation (\ref{generallangevin}) is
nonzero.
Furthermore, there is the case $\beta=1$.  As an illustration, in the
case $K=4$ there are two classes of special points for $\beta<1$. One
of them has $\omega_1=1/2$. In this case, the influence of the frozen
nodes will lead to $(N^{ini})^{4/5}$ nonfrozen nodes. Boolean
functions of the nodes with 2 nonfrozen inputs and the number of them
left after the first process are such that self-freezing loops are
made and their influence will again give $(N^{ini})^{2/3}$ as the
number of nonfrozen nodes in the network.  This case can obviously be
reduced to the $K=3$ network. The other class of special points is
obtained when the parameters of the network are $\omega_1=1/2$ and
$\omega_2=1/3$. In this case, $(N^{ini})^{6/7}$ nodes will be left nonfrozen
after determining the influence of the frozen nodes. One can easily
show that the creation of self-freezing loops is possible and that
their influence leads to a number of nonfrozen nodes that scales as
$(N^{ini})^{2/3}$ with the network size.

For general values of $K$, the $K-2$ classes of special points with
$\beta < 1$ are given by the condition $\omega_j=1/(j+1)$ for all
$j\in [1,i]$ where $i$ takes for every class one of the values from
the interval $[1,K-2]$. This means that
 $f_1=0,\dots,f_i=0$ in the Langevin equation (\ref{generallangevin})
and the term $f_{i+1}(\omega_1,\dots,\omega_{i+1})(N/N^{ini})^{i+1}$
is the leading one. The nodes left nonfrozen after determining the
influence of the nodes with constant functions scale with the network
size as
$(N^{ini})^{(2i+2)/(2i+3)}$. The numbers and Boolean functions of the
nodes with $k\in [2,i+1]$ nonfrozen inputs are such that they allow
for the creation of the self-freezing loops, and their influence
will for each of these special points, i.e., for each $i\in [1,K-2]$,
reduce the number of nonfrozen nodes to $(N^{ini})^{2/3}$.

For networks without constant functions (that is with $\beta=1$) the
frozen core arises only because of the creation of self-freezing loops
and their effect on the network. Just like for all other parameter
values, there is straightforward generalization of the analysis
performed for this type of networks in the case when $K=3$ in Section
\ref{noconstantfunctions}. In the case when $\omega_i=1/(i+1)$ for all
$i\in [1,K-1]$ there exists again a mapping of the self-freezing loops
on the relevant loops of a $K$ critical network with only reversible
and nonfrozen functions, from which it follows that the number of
nodes that are initially on self-freezing loops scales as
$(N^{ini})^{1/3}$. The process described in Section \ref{selffreezing}
can then be generalized to these networks. For any other choice of
parameters satisfying the criticality condition (\ref{gencritcond})
for $\beta=1$, self-freezing loops can also be formed, and after
moving only one of them in the container with frozen nodes we will
have the same process as for the one of the classes of critical
networks with $\beta<1$ that were already studied. Scaling of the
number of nonfrozen nodes in the critical networks without frozen
nodes and any fixed number of inputs will be the same as in all other
critical networks.

Let us end this section by noting that there is another class of
special points when the Boolean functions are chosen such that each of
them responds only to one of the $K$ inputs. In this case, the network
is effectively a $K=1$ network, since for each node those $K-1$ inputs
to which the node does not respond, can be cut off. In the
calculations of the previous sections we have always assumed that a
nonvanishing proportion of functions is not of this type.


\section{Relevant nodes and the number and length of attractors}
\label{relevant}

Relevant nodes are the nodes whose state is not constant and that
control at least one relevant node. These nodes determine completely
the number and period of attractors.  In the previous sections, we
have shown that the number of nonfrozen nodes scales as
$(N^{ini})^{2/3}$ for any critical network. We have also seen that
among them there are only $(N^{ini})^{1/3}$ nodes having two nonfrozen
inputs, and that the number of nonfrozen nodes with more than two
nonfrozen inputs vanishes in the thermodynamic limit.  The nonfrozen
nodes can now be connected to a network. This is a reduced network,
where all frozen nodes have been cut off. In
\cite{kaufmanandco:scaling}, we defined a stochastic process for
the formation of this reduced network and the identification of the
relevant nodes for critical $K=2$ networks.  The relevant nodes are
determined by removing iteratively nodes that are not relevant
because they influence only frozen and irrelevant nodes. The number
of relevant nodes was found to scale as $(N^{ini})^{1/3}$, and 
the scaling function characterizing their probability distribution
depends on the parameters of the model.

The scaling of the number of nonfrozen nodes as well as the scaling of
the number of nonfrozen nodes with two nonfrozen inputs as function of
the network size is the same for every critical network, as we have
shown in this paper. Since the fraction of nodes with more than two
nonfrozen inputs is vanishing in the thermodynamic limit, the network
of nonfrozen nodes, which is the starting point for the process of
determining the relevant nodes, is the same as in the $K=2$ case. So,
we can conclude that the results for the scaling of the number of
relevant nodes found in \cite{kaufmanandco:scaling} for the $K=2$
critical networks are valid for any critical network. The number of
relevant nodes in critical networks scales as $(N^{ini})^{1/3}$ with
the network size. Among them are a constant number of relevant nodes
with two relevant inputs and a vanishing number of relevant nodes with
more than two relevant inputs in the limit $N^{ini} \to \infty$. If
only these nodes and the links between them are considered, they form
loops with possibly additional links and chains of relevant nodes
within and between loops.

It follows that all critical networks with $K > 1$ show the same
scaling behavior. The only exception is the case $K=1$, which is
different because there is no frozen core.

As we have shown in \cite{kaufmanandco:scaling}, we can derive
properties of attractors from the results for the relevant nodes. In
particular, we can take over the result of
\cite{kaufmanandco:scaling} that all relevant components apart from
a finite number are simple loops, and that the mean number and length
of attractors increases faster than any power law with the network
size.

\section{Conclusions}
\label{conclusions}

In this paper, we have considered the limit of large network size, and
we have found the scaling behavior of the number of nonfrozen nodes,
of the number of nonfrozen nodes with more than one nonfrozen input,
of the number of relevant nodes, and of the number of relevant nodes
with more than one relevant input in a general class of critical
random Boolean networks with fixed number of inputs per node. The mean
values of these quantities scale with network size $N^{ini}$ as a
power law in $N^{ini}$, with the exponents being the same for any
critical network.  No matter what the distribution of the Boolean
functions is and how many inputs per node the critical network has,
number of nonfrozen nodes scales with the network size as
$(N^{ini})^{2/3}$, the number of nonfrozen nodes with two nonfrozen
inputs scales as $(N^{ini})^{1/3}$, the exponent for the number of
nonfrozen nodes with three nonfrozen inputs is zero, and it is $-n/3$
for the number of nonfrozen nodes with $n+3$ nonfrozen inputs. The
number of relevant nodes scales always as $(N^{ini})^{1/3}$,
with a constant number of them having two inputs and a vanishing
proportion having more than two. 

It follows that all critical random Boolean networks with $K>1$ belong
to the same class of systems. Changing the weights of the different
Boolean functions (for instance by choosing threshold networks or
canalyzing networks) or changing the number of inputs per node (which
might make the model more relevant for biological applications) will
not change the scaling of the number of nonfrozen and relevant nodes
with the size of the network, and it will not change the fact that the
number and length of attractors increases faster than any power law
with the network size, as long as the network is critical.  Using a
different method, Samuelsson and Socolar have recently also found that
the number of nonfrozen nodes scales in the same way for all $K>1$
critical networks \cite{samuelsson:exhaustive}.

From the calculations performed in this paper it can be concluded that
the results are also valid for networks that have nodes with different
values of $K$. If $K_{max}$ is the largest number of inputs occuring
in the network, we can set $K=K_{max}$, and we can view nodes with
less inputs as nodes with $K_{max}$ inputs, but with a function that
does not depend on all of its inputs. In contrast, our results cannot
be generalized to networks with a broad distribution of the number of
outputs. The method employed in this paper is based on a Poissonian
distribution of the number of outputs, and is most likely valid also
for other distributions as long as the second moment of the number of
outputs is finite. This can for instance be concluded from the analogy
between the propagation of activity in a Boolean network and
percolation on a directed graph, for which many results are known
\cite{schwartz:percolation}.

The finding that the number and length of attractors in critical
Boolean networks increases superpolynomially with network size is
detrimental to the hypothesis that these networks are models of gene
regulation networks, where only a limited number of dynamic pathways
should exist. However, by considering asynchronous update instead of
parallel update and by requiring that dynamics should be robust with
respect to fluctuations in the update sequence, the number of
attractors reduces to a power law in system size, which is more
realistic than the superpolynomial growth
\cite{klemm:asynchrstability,greil:dynamics}. The method presented in
this paper is independent of the updating scheme, and the scaling of
the number of nonfrozen and relevant nodes is therefore same for
asynchronous update as for parallel update. The relevant components
are consequently also the same.  With the insights obtained in the
present paper, we can immediately apply the results for asynchronous
update in $K=2$ critical networks to critical networks with larger
values of $K$, and we can conclude that the number of attractors in
critical networks with asynchronous update increases as a power law of
the system size.

Finally, let us consider networks where the connections between nodes
are not made at random, but that show some degree of clustering. Such
networks have a finite proportion of nodes that have correlated inputs
and that can therefore become frozen, e.g., because their inputs are
always in the same state. In contrast, the randomly wired networks
considered in the present paper have only a limited and small number of
nodes with correlated inputs even in the thermodynamic limit of
infinite network size.  For small-world networks, which have a high
degree of clustering, our method for determining the frozen core is
not valid, because it is based on the assumption that nodes choose
their inputs independently from each other. Small-world networks need
therefore a separate analytical treatment, which has not been done so
far.

Acknowledgements: we thank Viktor Kaufman for useful discussions.

\end{document}